# Superstatistics for fractional systems

## Gadjiev B. R.

International University for Nature, Society and Man "Dubna"

Abstract

The purpose of this paper is to develop a new fractional dynamical approach to superstatistics. Namely, we show that superstatistical distribution functions can be obtained from stationary solutions of the generalized Fokker-Planck equation for fractional systems by using the fractional generalization Bayes theorem. We present specific examples of such distribution functions for fractional systems.

**Introduction**

In principle, the probability distributions used in natural, biological, social and financial sciences can be formally derived by maximizing the entropy with adequate constraints [1]. The celebrated Shannon entropy, defined by

$$S = -\sum_i p_i \ln p_i \qquad (1)$$

or in a continuous approximation

$$S = -\int_0^\infty p(x) \ln p(x) dx, \qquad (2)$$

subjects to the normalization condition $\int_0^\infty p(x)dx = 1$. We consider the quantity

$$G_I = -\int_0^1 p \ln p \, dp, \qquad (3)$$

or in the discrete version $G_I = -\sum_i p_i \ln p_i \, \Delta p_i$.

Using fractional integrals we consider fractional generalizations of (3) which have the form

$$G_F = -\frac{1}{\Gamma(\alpha)} \int_0^1 \frac{1}{p^{1-\alpha}} p \ln p \, dp, \qquad (4)$$

Obviously, when $\alpha = 1$ we obtain the expression (3). Besides from expression (4) it is clear that

$$G_F = -\frac{1}{\Gamma(\alpha)} \int_0^1 p^\alpha \ln p \, dp. \qquad (5)$$

In the discrete version $\Gamma(\alpha) G_I = -\sum_i p_i^\alpha \ln p_i \, \Delta p_i$. We note that $G_{Shafee} = -\sum_i p_i^\alpha \ln p_i$ is the entropy of Shafee, which was introduced in [2]. When $\alpha=1$, an entropy of Shafee coincides with entropy of Shannon.

After the integration of the expression (5) over the $\alpha$ and dividing by the $(\alpha - 1)$ we obtain

$$\frac{1}{(\alpha-1)}\int_0^\alpha \Gamma(\alpha)G_F d\alpha = -\frac{1}{(\alpha-1)}\int_0^\alpha d\alpha \int_0^1 p^\alpha \ln p\, dp = -\frac{1}{(\alpha-1)}\int_0^1 (p^\alpha - 1)\, dp,$$

or in the discrete version at $\Delta p_i = 1$ we obtain

$$G_{Tsallis} = \sum_{i=1}^W \frac{p_i - p_i^\alpha}{\alpha - 1} \tag{6}$$

We note that $G_{Tsallis}$ is the entropy of Tsallis [3]. When α=1, an entropy of Tsallis coincides with entropy of Shannon. Therefore, the Chafee's entropy is connected with the fractal structure, whereas the Tsallis entropy corresponds to the multifractals.

**The fractional systems and superstatistics**

Complex systems in nature exhibit a rich structure of dynamics, described by a mixture of different stochastic processes on various time scales. Such dynamical processes with time scale separation are described in the framework of the superstatistical approach [4, 5].

In superstatistical approach "local equilibrium" is meant in a generalized sense for suitable observables of the system dynamics under consideration. In the long term, the stationary distribution of a superstatistical inhomogeneous system arises as superposition of a local factor $p(\varepsilon|\beta)$ with various values of $\beta$ weighted with a global probability density $g(\beta)$ to observe some value $\beta$ in a randomly chosen cell [4].

While on the time scale $T$ the local stationary distribution in each cell is $p(\varepsilon|\beta)$ the distribution describing the long-time behavior of the entire systems for $t \gg T$ defined by Bayes theorem

$$f(\epsilon) = \int_0^\infty d\beta\, g(\beta) p(\varepsilon|\beta), \tag{7}$$

and exhibits non trivial behavior [4].

For example, if $g(\beta)$ is $\chi^2$-distribution of degree $n$

$$g(\beta) = \frac{1}{\Gamma\left(\frac{1}{q-1}\right)\beta} \left[\frac{1}{(q-1)\beta_0}\right]^{\frac{1}{q-1}} \beta^{\frac{1}{q-1}-1} e^{-\frac{\beta}{\beta_0(q-1)}} \tag{8}$$

equation (7) generates Tsallis statistics, with nonextesivity index $q$ given by $q = 1 + \frac{2}{n+1}$:

$$p(u) = \frac{1}{\left(1 + (q-1)\beta_0 u^2\right)^{1/(q-1)}}. \tag{9}$$

To obtain this result, we used the well-known result

$$\int_0^\infty x^{v-1} e^{-\mu x} dx = \frac{1}{\mu^v} \Gamma(v) \tag{10}$$

Recently, the Fokker - Planck theory of superstatistics has been developed. It is shown that the Fokker-Planck equation admits then as a general stationary solution a superstatistical one. Three major universality classes often observed in nature, i.e., gamma, inverse gamma, and log-normal superstatistics, are discussed as special solutions [5].

The application of fractals in complex systems is far ranging, from the theory of finance to the theory of the dynamics of fluids in porous media [6]. The meaning of the fractional integration can be considered as integration in the same fractional space. A generalization of the

Fokker - Planck equation for the description of physical systems with a fractal phase space dedicated a series of works [6, 7].

Generalized Fokker - Planck equation for fractal systems is represented by the expression [6]

$$\frac{\partial \rho(x,t)}{\partial t} = \frac{\partial(a(x,t)\rho(x,t))}{\partial x^\alpha} - \frac{1}{2}\frac{\partial^2(b(x,t)\rho(x,t))}{\partial(x^\alpha)^2} \qquad (11)$$

Fractional generalizations of Bayes' theorem can be represented by an expression

$$p(\epsilon) = \frac{1}{\Gamma(\alpha)}\int_0^\infty \frac{d\mu}{\mu^{1-\alpha}} p(\epsilon|\mu) f(\mu) \qquad (12)$$

Note that when $\alpha = 1$, the Bayes theorem takes the standard form for systems with an isotropic phase space.

The specific form of the distribution function $p(\epsilon|\mu)$ and $f(\mu)$ can be defined as a solution of the corresponding Fokker-Planck equation for the random variables $\epsilon$ and $\mu$. More precisely, the generalized Fokker - Planck equation for fractal systems with distribution function $f = f(q_1, q_2, t)$ has the form

$$\frac{\partial f}{\partial t} = \sum_{k=1}^{n=2} \frac{\partial j_k}{\partial q_k^\alpha} \qquad (13)$$

where $j_k = j_k(q_1, q_2, t)$ is the current

$$j_k = j_k(q_1, q_2) = K_k f - \frac{1}{2}\sum_{l=1}^{2} Q_{kl}\frac{\partial f}{\partial q_l^\alpha}. \qquad (14)$$

Stationary distribution function $f = f(q_1, q_2)$ is the solution of the equation

$$\sum_{k=1}^{n=2} \frac{\partial j_k}{\partial q_k^\alpha} = 0 \qquad (15)$$

The stationary solution of the Fokker - Planck equation for fractal systems has the form [3]

$$\rho(x) = \frac{\rho_0}{b(x)} e^{2\int \frac{a(x)}{b(x)} dx^\alpha} \qquad (16)$$

where the coefficient $\rho_0$ is defined by the normalization condition. For example, if $a(x) = kx^\sigma$ and $b(x) = Dx^\tau$, then the stationary solution of the Fokker-Plank equation for fractal systems is the following distribution function

$$p(x) = p_0 x^{-\tau} e^{-\frac{2k\sigma x^{\alpha+\sigma-\tau}}{D}} \qquad (17)$$

We introduce the notation $\frac{2k\sigma}{D} = \mu_0 \mu^{\alpha+\theta}$ and for simplicity, in accordance with equation (15) we consider the fractal Fokker-Planck equation for the variable $\mu$ in the form

$$\frac{\partial(wg(\mu))}{\partial \mu^\alpha} - \frac{1}{2}\frac{\partial^2(qg(\mu))}{\partial(\mu^\alpha)^2} = 0 \qquad (18)$$

where $w = w_0 \mu^\theta$ and $q$ are constant. It is easy to show that the solution of this equation is

$$g(\mu) = g_0 e^{-\frac{2w_0 \alpha \mu^{\alpha+\theta}}{q}} \tag{19}$$

Using the distribution function (17) and (19) and the well-known integral

$$\int_0^\infty x^{v-1} e^{-\mu x^p} dx = \frac{1}{p} \mu^{-\frac{v}{p}} \Gamma\left(\frac{v}{p}\right) \tag{20}$$

we obtain the following expression for the distribution function of the entire system:

$$p(\epsilon) = \frac{f_0 p_0}{\Gamma(\alpha)} \int_0^\infty \frac{d\mu}{\mu^{1-(\alpha+\tau)}} e^{-\frac{2w_0 \alpha \mu^{\alpha+\theta}}{q}} e^{-\frac{1}{\mu_0} \mu^{\alpha+\theta} \epsilon^{\alpha+\sigma-\tau}} \tag{20}$$

In this case the variables $\epsilon$ and $\mu$ have the same time-scale, and for the distribution the entire system $p(\epsilon)$, we obtain the distribution function which is characteristic for complex systems:

$$p(\epsilon) = \frac{f_0 p_0}{\Gamma(\alpha)(\alpha+\theta)} \Gamma\left(\frac{\alpha+\tau}{\alpha+\theta}\right) \left(\frac{2w_0 \alpha}{q} + \frac{1}{\mu_0} \epsilon^{\alpha+\sigma-\tau}\right)^{-\frac{\alpha+\tau}{\alpha+\theta}} \tag{21}$$

**Discussion**

In conclusion, we note that in the present approach, we do not share the variables of the system into fast and slow variables. Fractional generalization of the Fokker - Planck equation and Bayes' formula allows us to obtain the distribution functions in the power law form. The presented approach can be easily extended to describe the distribution functions of complex networks, and we expect to use it to describe the properties of functional brain networks.